\documentclass[aps,prl,amsmath,amssymb,floatfix,twocolumn,amsmath,superscriptaddress,twocolumn,nofootinbib,tighten,letterpaper]{revtex4-2}
\usepackage{multirow}
\usepackage{bbold}
\usepackage{subfigure}
\usepackage{color}
\usepackage{mathrsfs}
\pdfoutput=1 
\usepackage{amsfonts}
\usepackage{amsmath,bm} 
\usepackage{graphics} 
\usepackage{graphicx}
\usepackage[]{epsfig}
\usepackage{amsfonts}
\usepackage{subfigure}
\usepackage{bbold}
\usepackage{slashed}

\def\ba{\begin{eqnarray}}
\def\ea{\end{eqnarray}}
\def\be{\begin{equation}}
\def\ee{\end{equation}}

\begin{document}
\title{Engineering holographic flat fermionic bands
}

\author{Nicol\'as Grandi}
\email{grandi@fisica.unlp.edu.ar}
\affiliation{Instituto de F\'\i sica de La Plata - CONICET, C.C. 67, 1900 La Plata, Argentina.}
\affiliation{Departamento de F\'\i sica - UNLP, Calle 49 y 115 s/n, 1900 La Plata, Argentina.}

\author{Vladimir Juri\v ci\' c}
\email{juricic@gmail.com}
\affiliation{Departamento de F\'isica, Universidad T\'ecnica Federico Santa Mar\'ia, Casilla 110, Valpara\'iso, Chile.}
\affiliation{Nordita, KTH Royal Institute of Technology and Stockholm University,
Hannes Alfvéns väg 12, 106 91 Stockholm, Sweden}

\author{Ignacio Salazar Landea}
\email{peznacho@gmail.com}
\affiliation{Instituto de F\'\i sica de La Plata - CONICET, C.C. 67, 1900 La Plata, Argentina.}

\author{Rodrigo Soto-Garrido}
\email{rodrigo.sotog@gmail.com}
\affiliation{Facultad de F\'isica, Pontificia Universidad Cat\'olica de Chile, Vicu\~{n}a Mackenna 4860, Santiago, Chile}
\begin{abstract}
{ In electronic systems with flat bands, such as twisted bilayer graphene, interaction effects govern the structure of the phase diagram. In this paper, we show that a strongly interacting system featuring fermionic flat bands can be engineered using the holographic duality.
In particular, we find that  in the holographic nematic phase, two bulk Dirac cones separated in momentum space at low temperature, approach each other as the temperature increases. They eventually collide at a critical temperature yielding a flattened band with a quadratic dispersion. On the other hand, in the symmetric (Lifshitz) phase, this quadratic dispersion relation holds for any finite temperature. 
We therefore obtain a first holographic, strong-coupling realization of a topological phase transition where two Berry monopoles of charge one merge into a single one with charge two, which may be relevant for  two- and three-dimensional topological semimetals. 

}
\end{abstract}
\maketitle
%
\noindent\emph{Introduction.} Flat electronic bands have recently attracted  significant attention due to the experimental realization of twisted bilayer graphene \cite{cao2018correlated,cao2018unconventional}. They  naturally promote interaction effects as dominant, leading to a rich landscape of possible strongly correlated phases \cite{andrei2020graphene}. Strongly interacting systems, however, may be prohibitively difficult to address within the traditional frameworks, such as perturbation theory. Furthermore, numerical methods are restricted to exact diagonalization due to the sign problem present in systems involving fermionic degrees of freedom.

Given the necessity of using a nonperturbative approach, the AdS/CFT correspondence, also known as  holographic duality \cite{Maldacena:1997re,Witten-1998,Ammon:2015wua}, has been broadly used to investigate strongly coupled systems. In the more general sense, this duality proposes an equivalence between systems described by strongly interacting quantum field theories and weakly coupled systems in a curved spacetime with an extra spatial dimension. As such, these holographic methods so far have been useful to gain qualitative insights on condensed matter systems, for instance,  non-Fermi liquids, high T$_c$ superconductors, and topological systems, among others 
\cite{Zaanen:2021,hartnoll2018,Zaanen-2015}. Given that flat bands favor the effects of the strong electronic interactions, including the ensuing new quantum phases and phase transitions, it is then rather natural to apply the holographic duality to address the physics at strong coupling therein.


In this context, it is  worthwhile emphasizing that a Lifshitz geometry may be the natural setting for the construction of the holographic flat bands, given that, at least from a perturbative point of view,  the scaling of the  electronic density of states [$\rho(\varepsilon)$] with energy ($\varepsilon$) in $d$ spatial dimensions is tunable by the dynamical exponent ($z$), $\rho(\varepsilon)\sim|\varepsilon|^{\frac{d}{z}-1}$. It is then expected that a tunable dynamical exponent could  yield an instability towards a symmetry broken phase, as in the model for multi-Weyl semimetal we have recently proposed~\cite{Grandi:2021bsp}, realizing a phase transition into a nematic phase, as also found in Ref.~\cite{Dantas:2019rgp}. 
However, none of these holographic models feature the fermions in the bulk geometry that may in turn explicitly yield the holographic flat fermionic bands. 

Here, we address this important problem by incorporating the fermions directly into the model of Ref. \cite{Grandi:2021bsp}. By computing the Green function in the background geometry, we show  that the  dispersion relation of the fermions indeed displays the flattening feature. 
In particular, we find that  in the nematic phase,   
two Dirac cones separated in momentum space at low temperature, approach each other as the temperature increases. Eventually, at the critical temperature, they collide yielding a flattened band with a quadratic dispersion relation. 
On the other hand, in the symmetric (Lifshitz) phase, this quadratic dispersion relation holds for any finite temperature. Therefore, besides the holographic band flattening, we find a first holographic, strong-coupling realization of a topological phase transition where two Berry monopoles of charge one merge into a single monopole with charge two, which pertains to  topological metals in both two \cite{wunsch2008dirac,montambaux2009merging,Tarruell:2012zz} and three \cite{Zhang2017NatPhys,RGJ2017,BelopolskiarXiv2021,GGLiuarXiv2021}  spatial dimensions. 


\vspace{2mm}
\noindent\emph{Free model.} Let us first discuss the flat bands within a $(2+1)$-dimensional free Dirac fermion model. Our aim is to extract from it the symmetry breaking pattern that we will later export into the holographic realm. It is defined by the Hamiltonian
\begin{equation}
    H_D'\!=\! -\gamma^t(\gamma^x p_x+\gamma^y p_y)\otimes \mathbb{1}_{\mbox{\scriptsize$2\!\times\!2$\normalsize}}+ i\,{m_*} \!\left(\gamma^x \!\otimes\! \sigma_2 -\gamma^y \!\otimes\! \sigma_1  \right),
\label{H}
\end{equation}
where $\gamma^\mu= (\sigma_3,-i\sigma_2,i\sigma_1)$ are the Dirac gamma matrices, and the Pauli matrices $\sigma_i$ in the second term act on an internal flavour index.
%
This 
is analogous to 
the construction of the effective 
Hamiltonian for bilayer graphene by coupling two single layers each containing a linearly dispersing Dirac fermion. 
%
%
The
spectrum reads
\begin{equation}
    \omega=\pm m_* \pm \sqrt{p_x^2+p_y^2+m_*^2}
\end{equation}
with gapped conduction and valence bands corresponding to both positive or both negative signs respectively, while otherwise the valence and conduction bands cross at zero energy. In the latter case, 
we obtain a 
low-energy ($p_x^2+p_y^2\ll m_*^2$)
quadratic dispersion 
\begin{equation}
    \omega\approx \pm \frac 1{2m_*}\left(p_x^2+p_y^2\right),
\end{equation}
which still preserves rotational invariance. 

The wave equation resulting from the Hamiltonian in Eq.~\eqref{H} can be obtained from the action
\begin{equation}
    S=S_{\sf free}-i\int d^3x\, \bar\Psi \slashed{W}\Psi.
    \label{S}
\end{equation}
Here the first term is the free Dirac action for a pair of two-component spinors, while the deformation corresponds to a coupling of the pair to a constant non-Abelian vector field $W=m_*(\sigma^{1}dx+\sigma^{2}dy)$. This explicitly breaks the $U(2)$ symmetry down to $U(1)$.  Notice that spatial rotational invariance is also broken. However,  there is a ``mixed'' rotational symmetry preserved in the system which is realized by compensating a spatial rotation with an internal transformation generated by $\sigma_3$. The latter is enough to preserve the rotational invariance of the spectrum.

\vspace{2mm}
\noindent\emph{Holographic model.}
We now construct a bottom up holographic dual to a strongly coupled version of the model in Eq.~ \eqref{S}. To do so,
we first extend the global boundary symmetries to the bulk as gauge symmetries, introducing the relevant bulk gauge fields. Gauged space-time symmetries require a dynamical metric, minimally described by Einstein gravity. We include a negative cosmological constant in order to get AdS asymptotics. Moreover, we add the corresponding Yang-Mills fields needed to gauge the boundary $U(2)$ symmetry  
\begin{equation}
S= 
\int d^4x\sqrt{-g}\left(R-2\Lambda\right)
-\frac{1}{4}\int
\left[ F\wedge {^{\!\star}  F} +
\mathrm{Tr}\left(
G\wedge{^{\!\star} G}\right)
\right]\,,\label{eq:HoloAct}
\end{equation}
where ${F}=d A$ represents a $U(1)$ gauge field strength accounting for the conservation of particle number, while $G=d{B} - i(q/2) {B}\wedge B$ is the strength of a $SU(2)$ gauge field, accounting for the flavor symmetry in UV of the model in Eq.~\eqref{S}. One may therefore roughly state that, at this step, matter is hidden behind the black hole horizon, and an exterior observer can only see the total charges.

Hence, we look for asymptotically AdS black holes solutions, with the generic \emph{ansatz} for the metric
\begin{equation}
\mathrm d s^2 = \frac{1}{r^2}\Big(\!-\!N f\mathrm dt^2 + \frac{dr^2}{f}+ dx^2 + dy^2+ 
2 h\, dx\, dy \Big)\,,
\label{eq:metrica}
\end{equation} 
in terms of purely $r-$dependent functions $f,N$ and $h$, which close to the boundary satisfy $f,N\to 1$ and $h\to  0$. To describe a boundary system at finite temperature, we focus on black hole solutions with a horizon at finite $r=r_h$, at which $N$ and $h$ are bounded, and $f$ vanishes linearly with $f'= {4 \pi T}/{\sqrt{N}}$.

For the gauge fields, we write
\begin{align}
~\label{eq:ansatz}
  \,\,\, {A}&=0\\
  B &=  \frac12\left(Q_{1} \sigma_1\,+Q_{2} \sigma_2\,\right) dx +  \frac12\left(Q_{1}\sigma_2+Q_{2}\sigma_1 \right)dy
\nonumber \,.
\end{align}
The gauge field $A$,  coupled at the boundary to the particle current, is turned off, implying that the chemical potential vanishes. Here $Q_{1,2}$ are $r$-dependent functions which are finite at the horizon. On the boundary, we turn on a deformation  analogous to the second term of the action~\eqref{S}, by requiring $B\to W$ or equivalently $Q_1\to 2m_*$.  This reproduces the symmetry breaking pattern of our free model in the present holographic ({\em i.e.} strongly coupled) setup. This approach has already been used to construct holographic duals of fermions with particular dispersion relations, as, for instance, Weyl semimetals \cite{Landsteiner:2015lsa}, multi-Weyl semimetals \cite{Dantas:2019rgp,Juricic:2020sgg}, nodal line semimetals \cite{Liu:2018bye}, and Weyl $Z_2$ semimetals \cite{Ji:2021aan}. 
 
\begin{figure}[t]
    \centering
    \includegraphics[width=0.47\textwidth]{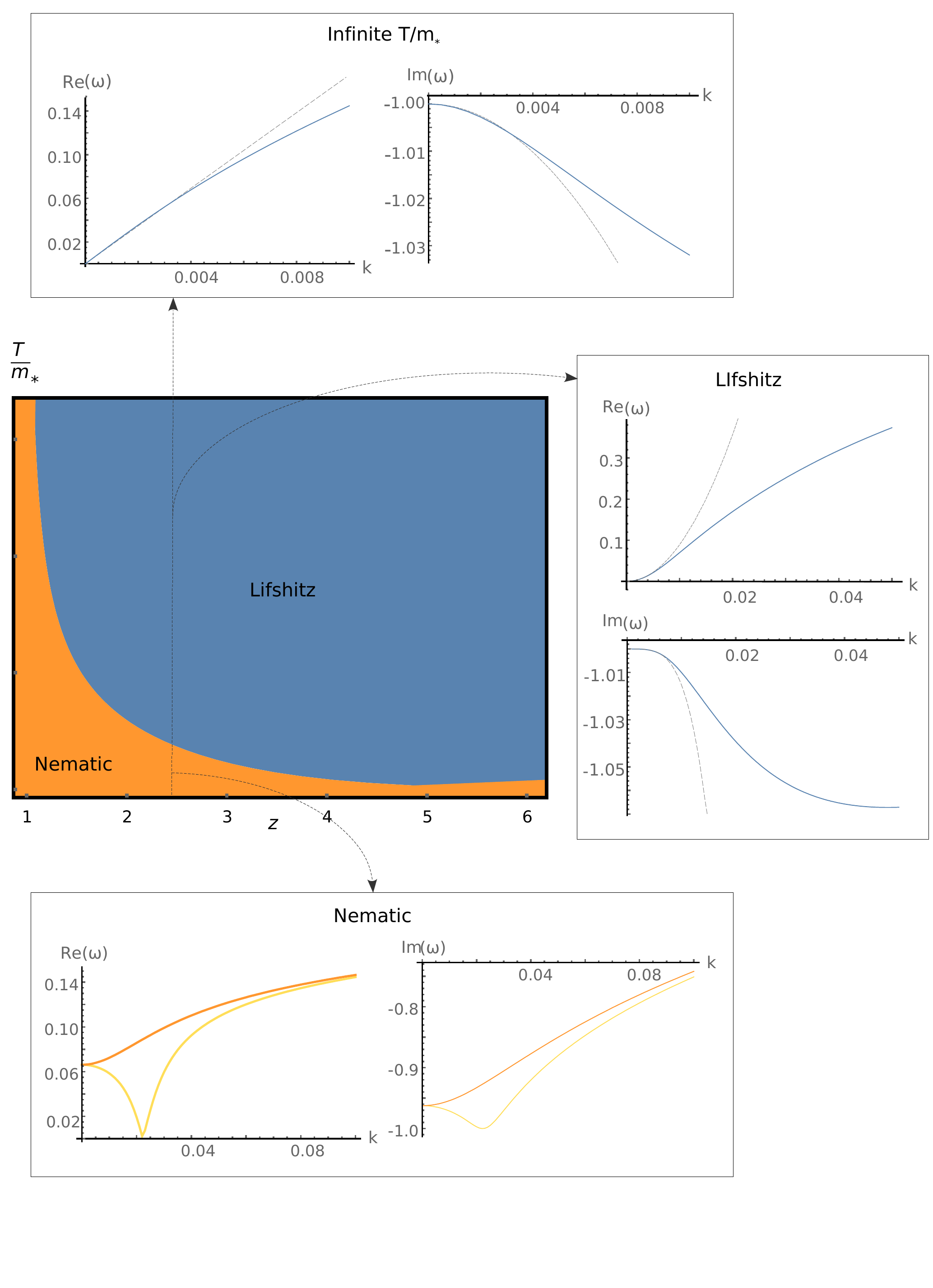}
    \vspace{-.75cm}
    \caption{Phase diagram of the model in Eq.~\eqref{S}, showing the nematic phase close to the origin and the Lifshitz one away from it. At the top, we see the fermionic dispersion relation for infinite $T/m_*$, the dotted lines corresponding to the approximation in Eq.~\eqref{relf}. On the right hand side, the dispersion corresponding to the Lifshitz phase for $T/m_*=47.74$, with the lines of approximation \eqref{flatf} is displayed. At the bottom, it is shown the anisotropic dispersion relation in the nematic phase, just below the critical temperature $T/m_*\approx0.075$, along the two directions of Fig.\ \ref{cones}. We here set $q=q_D= 2$, {\em i.e.} $z=2.48$.}
    \label{Jxyplot}
\end{figure}

The resulting phase diagram  in terms of the gauge coupling $q$ and the dimensionless temperature $T/m_*$ is shown in Fig.\ \ref{Jxyplot}.  
At high enough $T/m_*$ and $q$, only  rotational invariant solutions are present, with $h$ and $Q_2$ equal  exactly zero. The model flows towards a metastable Lifshitz geometry in the IR, with a scaling exponent $z$ that is completely fixed by the coupling $q$ as the root of  a cubic polynomial \cite{Devecioglu:2014iia}
\begin{equation}\label{eq:zq}
    q^2(z^3+z^2)+(q^2-24)z-3(q^2+8)=0\,.
\end{equation}
As we lower the temperature or the gauge coupling, a boundary nematic phase develops, since the fields $Q_2$ and $h$ in the bulk  spontaneously acquire non-vanishing expectation values. Full $z=1$ rotational invariance is however recovered in the deep IR. 

\vspace{2mm}
\noindent\emph{Fermions in the bulk.} To confirm that our holographic model actually describes flat bands, we need to obtain its fermionic dispersion relation. In holography, boundary matter fluctuations are represented by the fermionic degrees of freedom $\Psi$ in the bulk, satisfying the Dirac equation  \cite{Henningson:1998cd} 
\be
\Gamma^A \, D_A\Psi =0\,.
\label{eq:eom-fermionic}
\ee
Here $\Gamma^A$ with $A\in\{0,1,2,3\}$ are the $(3+1)$-dimensional flat space Dirac gamma matrices, and $D_A$ is the curved space and gauge covariant derivative. The field $\Psi$ is an $SU(2)$ doublet of $(3+1)-$dimensional spinors, containing a total of eight independent components. 

As our model is closely related to the non-Abelian $p$-wave superconductor  
\cite{Gubser:2008wv,Ammon:2008fc,Ammon:2009xh,Arias:2012py, Gubser:2010dm,Ammon:2010pg}, here we only sketch the construction of $D^A$.
The key step is to write the metric in the tetrad formalism 
\be
ds^2 = 
\eta_{AB}e^Ae^B\,,
\ee
where $\eta_{AB}=\text{diag}(-1,1,1,1)$ is the Minkowski metric and $e^A
$ are the coframe fields, given by
\begin{align}
&e^1 \equiv \displaystyle\frac{1}{2r}\left[\left( \sqrt{1\!+\!h}+\sqrt{1\!-\!h}\right)dx+\left( \sqrt{1\!+\!h}-\sqrt{1\!-\!h}\right)dy\right]\,,\nonumber\\
&e^2 \equiv \displaystyle\frac{1}{2r}\left[\left( \sqrt{1\!+\!h}-\sqrt{1\!-\!h}\right)dx+\left( \sqrt{1\!+\!h}+\sqrt{1\!-\!h}\right)dy\right]\,,\nonumber\\
& e^3 \equiv \displaystyle\frac{1}{r\sqrt{f}}\;dr \,,
\qquad\quad
e^0 \equiv \displaystyle\frac{1}{r}\sqrt{N f }\;dt\,.
\label{eq:vielbein}
\end{align}
The corresponding spin connection 
can be then found through the torsionless condition, for details and conventions see \cite{Giordano:2016tws}.

We furthermore conveniently perform  Fourier transform  from the transverse coordinates  $(t,x,y)$ to $(\omega,k_x,k_y)$, and re-scale the spinors as 
\be 
\Psi  ={r^{3/2}}(fN(1-h^2))^{-1/4} \psi_k(r)\,.
\ee
Projecting into the $2+1$ transverse spacetime by considering eigenvectors of the radial gamma matrix
\be
\psi_k=\left(\begin{array}{c}
\psi_+\\
\psi_- 
\end{array}\right)\,,
\ee
we obtain the explicit form of the equations for the two components $\psi_\pm$,

\begin{align}
&\psi'_-- \frac{i}{\sqrt{2f(1-h^2)\tilde h}}U\cdot \psi_+ =0\,,\nonumber\\
&\psi'_++ \frac{i}{\sqrt{2f(1-h^2)\tilde h}}U\cdot \psi_- =0\,,
\end{align}
with  $\tilde h\equiv 1+\sqrt{1-h^2} $, $H\equiv h+i\tilde h$ and
\begin{widetext}
\begin{equation}
U=\begin{pmatrix}
 \tilde h k_y- h k_x  & \tilde h k_x- h k_y+\sqrt{\frac{2(1-h^2)\tilde h }{fN}}\omega& \frac12 q_D \left(HQ_1-iH^*Q_2\right) & \frac12 q_D \left(HQ_2-H^*Q_1\right)\\
\tilde h k_x- h k_y+\sqrt{\frac{2(1-h^2)\tilde h }{fN}}\omega & h k_x- \tilde h k_y   & \frac12 q_D \left(HQ_2-iH^*Q_1\right) & \frac12 q_D \left(H^*Q_1+iH^* Q_2\right) \\
 \frac12 q_D \left(H^*Q_1+iHQ_2\right) &  \frac12 q_D \left(-iH^*Q_1-H^*Q_2\right) & \tilde h k_y- h k_x & \tilde h k_x- h k_y+\sqrt{\frac{2(1-h^2)\tilde h }{fN}}\omega\\
\frac12 q_D \left(-i H Q_1+H^* Q_2\right) & \frac12 q_D \left(-H^*Q_1-iHQ_2\right) & \tilde h k_x- h k_y-\sqrt{\frac{2(1-h^2)\tilde h }{fN}} & hk_x-\tilde h k_y
\end{pmatrix}.
\end{equation}
\end{widetext}

Close to the boundary, the fields $\psi_\pm$ tend to a constant $\psi_\pm\to \psi_\pm^{UV}$. The four components of $\psi_+^{UV}$ are interpreted as the expectation values of the dual fermionic operator, while those of $\psi_-^{UV}$ are instead proportional to the external sources. On the other hand, close to the black hole horizon, in-going boundary conditions yield
\begin{equation}
\psi_\pm=\psi_{h\pm}(r-r_h)^{-\frac{i\omega}{4\pi T}}
\label{ingoing}
\end{equation}
with  $\psi_{h+}=-i (I_{2\times2}\otimes\sigma_2)\cdot\psi_{h-}$.
This constraint implies that $\psi_\pm^{UV}$ are not all independent. Indeed, since the equations are linear, we can assume the linear relations $\psi_+^{UV}=M_+\psi_h$ and $\psi_-^{UV}=M_-\psi_h$, where the matrices $M_\pm$ must be obtained by numerical integration. This in turn implies $\psi_+^{UV}=M_+M_-^{-1}\psi_-^{UV}$, allowing us to identify the retarded fermionic correlator as $G=M_+M_-^{-1}$. 
Then the poles of the correlator can be simply read off from the zeros of the determinant of $M_-$.

\vspace{2mm}
\noindent\emph{Fermionic dispersion relation.} 
%
Focusing on the lowest lying poles of 
the retarded correlator, we obtain the dispersion relation of the holographic fermion. Starting at the infinite temperature limit, 
its form reads 
\begin{equation}
    \omega \approx -i+ v_f k-i\Gamma k^2 +\dots
    \label{relf}
\end{equation}
featuring  a purely imaginary gap and a real part scaling 
linearly with $k$, as expected for a massless fermion. At large $k$  non-linear contributions coming from the broken conformal symmetry have to be included. The corresponding dispersions are depicted in the plots at the top of Fig.\ \ref{Jxyplot}.

As  the temperature is lowered,
the fermionic dispersion relation \eqref{relf} gets modified to
\begin{equation}
    \omega \approx -i+ c k^2-i\Sigma k^4 +\dots
    \label{flatf}
\end{equation}
in agreement with expectations based on the free model.
%
This can be seen in the plots at the center-right of Fig.\ \ref{Jxyplot}
We emphasize that the dispersion relation in Eq.~\eqref{flatf}  fits to a quadratic real part close enough to the origin.
This behavior holds until the system enters the instability towards the nematic phase. The dispersion relation then  becomes direction dependent. To illustrate this behavior,  two representative cuts on the ${\bf k}$ plane are shown at the bottom plots of Fig.\ \ref{Jxyplot}.
We observe a rather notable breaking of the rotational invariance. 
Interestingly,
there is
a zero frequency excitation at some finite $k=k_*$ along one of the 
directions, close to which
a relativistic behavior 
shows up. 
${\sf Re}\,\omega\approx v_*(k-k_*)$. This 
is a consequence of
a $z=1$ fixed point, which corresponds to the deep IR of the zero temperature limit in the nematic phase.

\begin{figure}[h!]
    \includegraphics[width=0.47\textwidth]{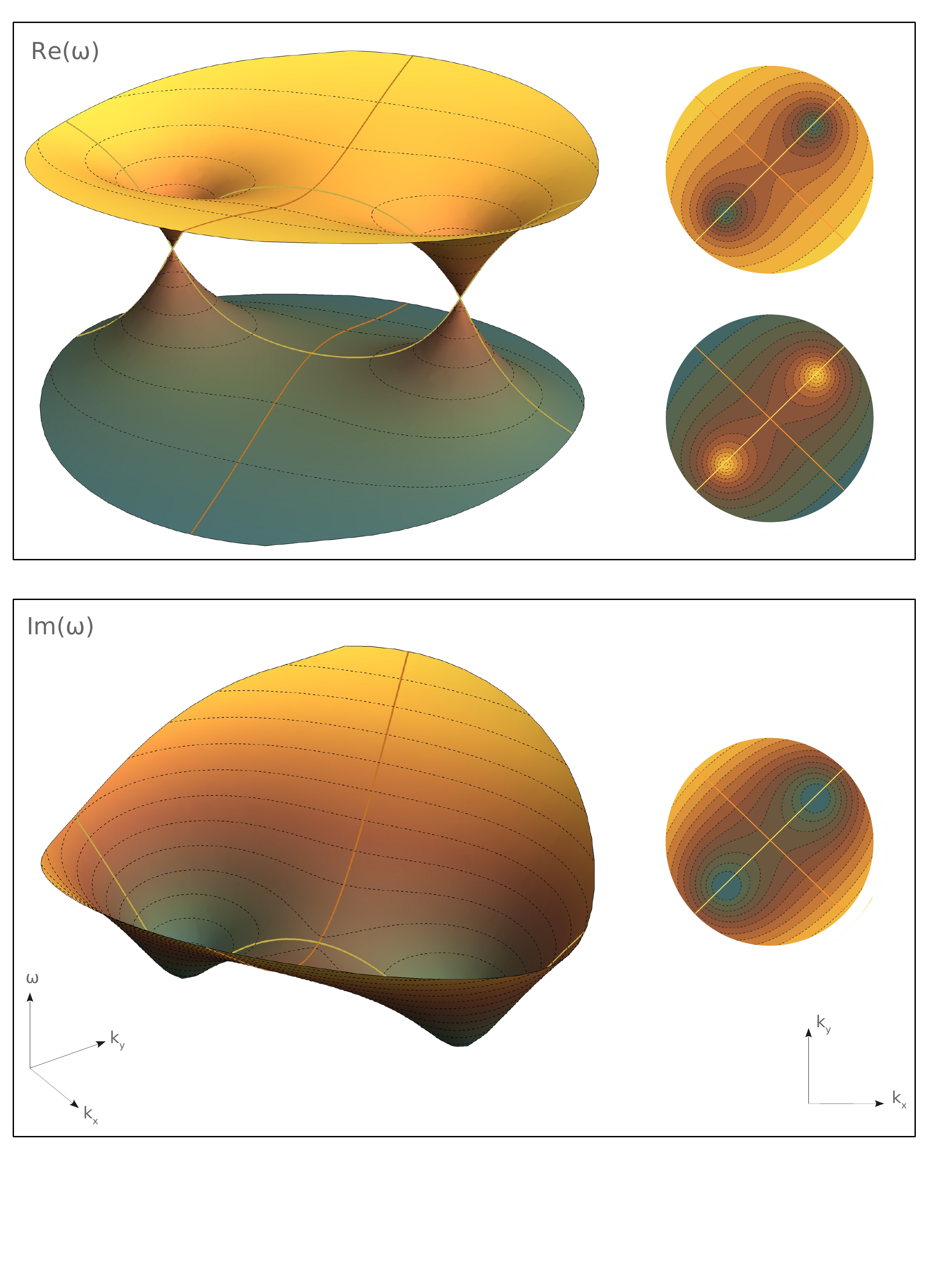} 
\vspace{-1.1cm}    
\caption{ Real (top) and imaginary (bottom) parts of the quasinormal frequencies  as a function of $\vec k= (k_x,k_y)$ in the nematic phase, just below the critical temperature $T/m_*\approx0.075$. The contour plots on the right hand side correspond to the 3D plots on the left. The orange and yellow lines denote the cuts which were plotted at the bottom of Fig.\ \ref{Jxyplot}.}
    \label{cones}
\end{figure}

Additionally, in Fig.~\ref{cones} we show a 3D plot with the position of the poles of the two point function in the momentum space. We see explicitly that rotational symmetry is broken while a discrete $Z_2$ symmetry around the $k_x+k_y$ axis is preserved.

\begin{figure}[t!]
    \centering 
    \includegraphics[width=0.48\textwidth]{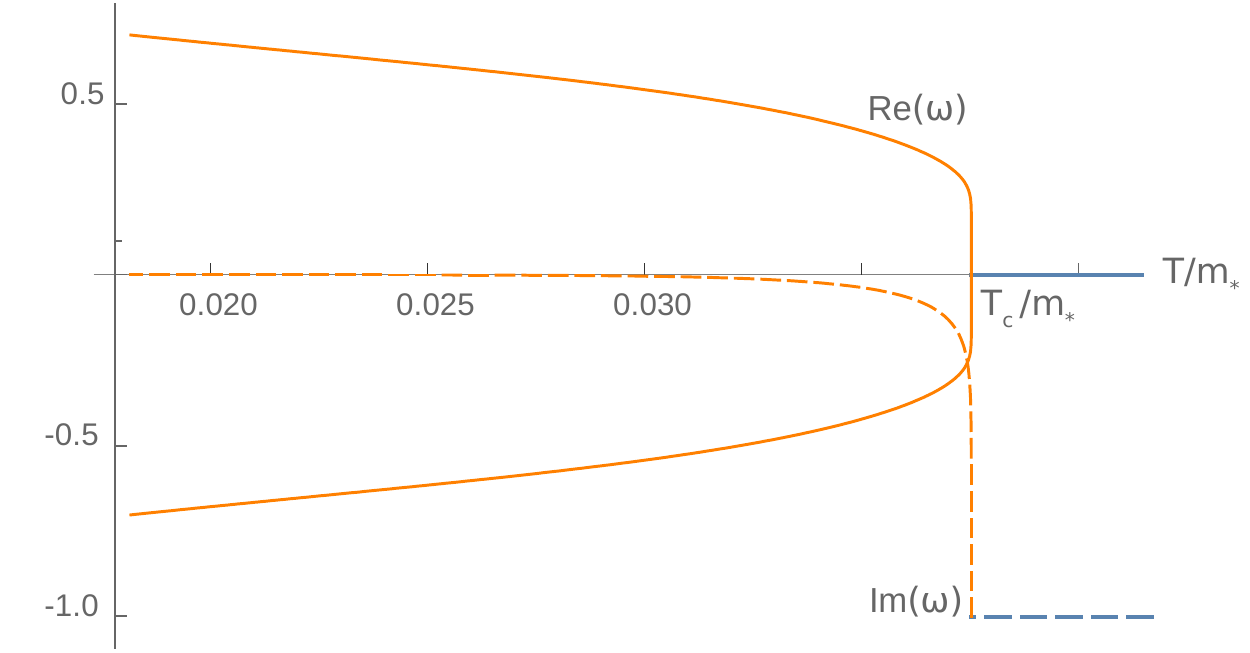}
    \caption{Real and imaginary part of the quasinormal frequencies at ${\bf k}=0$ as a function of the  temperature for the nematic phase.}
    \label{nemo}
\end{figure}

As we can see in Fig.\ \ref{nemo} the real part of the zero momentum quasinormal frequency quickly grows for the nematic phase, while the imaginary part drops to zero. These gaps lift the degeneracy at low energies according to the intuition that flat bands in interacting systems should be generically unstable. On the other hand, the imaginary part drops to zero, making the quasiparticles long lived as the temperature is lowered.

\vspace{2mm}
\noindent\emph{Conclusions and outlook.}
In this paper, we realized a construction of fermionic flat bands in the context of holography by directly including the fermions in the bulk geometry within a standard bottom-up approach. We consider a holographic dual of a free fermionic model that mimics two copies of Dirac fermions hybridizing as in a graphene bilayer and yielding a quadratic low-energy dispersion. Interestingly, we found that the holographic  fermions inherit this quadratic dispersion relation from the free model, while the thermodynamics of the system points towards a different Lifshitz exponent. This should not be surprising given that the holographic system is strongly coupled and the critical exponents are indeed expected to  renormalize.




We here point out that since a  holographic system is intrinsically strongly coupled, and, on the other hand, a flat electronic band is highly  degenerate, we  expect that the holograhic flat band undergoes a phase transition towards a phase where the degeneracy is lifted. This is in fact consistent with the present holographic setup, since the  Yang-Mills black holes, featured in the model, are generically unstable. This fact is then reflected as a phase transition towards a nematic phase. The concomitant rotational symmetry breaking then leads to the   splitting of the quadratically dispersing nodes into two relativistic Dirac points that move apart as we lower the temperature. As such, this process can be thought as a holographic realization of the Berry monopole splitting at strong coupling.  


We will conclude by discussing some future directions. First, one may consider
a UV completion of our bottom-up model where the relation between the bosonic and fermionic sector arises from a stringy construction \cite{Davis:2011gi, Grignani:2014vaa,Ammon:2009fe,Fadafan:2020fod}. Furthermore,  the free fermionic construction can be extended to an arbitrary integer $n$ with the dispersion $\omega \sim k^n$ by using $n$ fermionic flavors and the generators of the spin-$(n-1)/2$ representation of the $SU(2)$ group \cite{Dantas:2019rgp}. It would be therefore interesting to explore holographic duals of the models with higher-$n$ fermionic dispersions.
Finally, one may consider generalizations of these fermionic models to non-relativistic Goldstone bosons \cite{Amado:2013xya,Schafer:2001bq}.

\vspace{2mm}
\noindent\emph{Acknowledgments:}  This work was supported by ANID-SCIA-ANILLO ACT210100 (R.S.-G. and V.J), the Swedish Research Council Grant No. VR 2019-04735 (V.J.), Fondecyt (Chile) Grant No. 1200399 (R.S.-G.), the CONICET grants PIP-2017-1109 and PUE 084 ``B\'usqueda de Nueva F\'isica”, and by UNLP grants PID-X791 (I.S.L. and N.E.G.).

 \bibliographystyle{apsrev4-1} \bibliography{references}

\end{document}